\documentclass
[preprint,showpacs,byrevtex,prl,10pt,a4paper,twocolumn,tightenlines]{revtex4}%
\usepackage{amsfonts}
\usepackage{amsmath}
\usepackage{amssymb}
\usepackage[dvips]{graphicx}%
\begin{document}
\preprint{ }
\title{Slow conductance relaxations; Distinguishing the Electron Glass from extrinsic mechanisms}
\author{Z. Ovadyahu}
\affiliation{Racah Institute of Physics, The Hebrew University, Jerusalem 91904, Israel }

\begin{abstract}
Slow conductance relaxations are observable in a many condensed matter
systems. These are sometimes described as manifestations of a glassy phase.
The underlying mechanisms responsible for the slow dynamics are often due to
structural changes which modify the potential landscape experienced by the
charge-carriers and thus are reflected in the conductance. Sluggish
conductance dynamics may however originate from the interplay between
electron-electron interactions and quenched disorder. Examples for both
scenarios and the experimental features that should help to distinguish
between them are shown and discussed. In particular, it is suggested that the
`memory-dip' observable through field-effect measurements is a characteristic
signature of the inherent electron-glass provided it obeys certain conditions.

\end{abstract}
\maketitle

\section{Introduction}

Many phenomena in solid state systems involve slow conductance changes. The
phenomenon that has been most studied in this context is conductance noise,
which often exhibits a 1/f power-spectrum. In degenerate Fermi systems
(metals, heavily-doped semiconductors) the prevalent view is that the
fluctuations in the conductance $G$ reflect temporal changes in the potential
experienced by the charge carriers \cite{1}. Such potential fluctuations may
be structural, involving slow dynamics of ions/atoms which, in turn, may be
triggered by a modified state of local charge following, for example,
electronic re-arrangement of valence electrons. These as well as "purely"
ionic displacement are referred to in this paper as structural
two-level-systems. Slow release/trapping of carriers will likewise manifest
itself in slow conductance fluctuations. Either mechanism may lead to
conductance fluctuations that typically extend to very low frequencies
\cite{1}.

As a rule, 1/f experiments were performed in systems where the average $G$ was
time independent, namely, in equilibrium or near equilibrium situations. There
are however many cases where $G$ itself changes slowly with time, which, by
definition suggests a non-equilibrium phenomenon. This may occur for example
due to annealing of defects, neutron irradiation, diffusion of an injected
dopant, illumination by light, and many other instances involving changes of
the potential landscape, or the density of carriers in the conducting system.
It often happens in these cases that the sluggish response observed in $G$
exhibits features that are characteristic of glasses. Experimental results
illustrating two mechanisms for such `extrinsic' glassy effects in
conductivity will be shown and discussed in section III.

Glassy effects in $G$ may arise from an \textit{intrinsic} mechanism, in which
case both the ions potential and the carriers density may be, in theory, time
independent. Glassy behavior of the electronic system in a system with
quenched disorder has been anticipated by several authors \cite{2,3,4,5}. This
new type of a non-ergodic system was first termed Electron Glass by Davies
\textit{et al} referring to a system with localized electronic states
interacting via an un-screened Coulomb interaction \cite{4}. Glassy effects
inherent to the electronic systems that may arise even in the non-interacting
system were also considered \cite{3}. However, common to all models of
intrinsic electron glass is a sufficiently strong static disorder such that
the electronic states are Anderson localized. This is one of the essential
differences between the electron glass and the extrinsic mechanisms alluded to
above. In the latter, glassy effects may manifest themselves even when the
transport mode is diffusive. By contrast, a \textit{pre-requisite} for
intrinsic electron glass behavior is that the system is Anderson localized,
and thus charge transport must be activated. This condition is necessary but
not sufficient, more detailed criteria will be given in the discussion section below.

Non-ergodic effects such as slow relaxation, aging, and other memory effects
associated with electron glass behavior were reported in few systems
\cite{6,7,8,9,10,11}. Naturally, the question of extrinsic effects dominating,
or at least, contributing to the measured effects have been considered in
early publications \cite{12}.

There are difficulties in categorically ruling out extrinsic effects;
Essentially all glasses, whether mechanical, magnetic, or electronic, show
similar dynamical features. While these universal attributes give the impetus
to study glasses, the similar effects make it hard to distinguish between
different glasses, especially when both types are revealed through the
\textit{same} measurable (\textit{i.e.}, conductance).\ Current theories of
the electron glass are not sufficiently detailed to allow this
differentiation. The one issue on which theory predicts a qualitative feature
that, experimentally, seems to be peculiar to the electron glass, is the law
of relaxation being logarithmic in time \cite{13}. This has indeed been
verified in several electron glasses \cite{8}, and in $In_{2}O_{3-x}$ this
relaxation law has been observed over almost 6 decades in time \cite{14}. It
was also shown in this system that the ions/atoms dynamics is associated with
quite different time scales and spatial scales than the dynamics of the
electron glass \cite{15}. Similar results were obtained on granular al films
by Grenet and by Delahaye \textit{et al} \cite{16}. However, the most
compelling evidence for an inherent electron glass is the dramatic dependence
of the glassy properties on the system carrier concentration \cite{17}. This
will be further elucidated in this paper.

In some cases it is possible to identify the origin of the glassy effects in
$G$ as arising from an extrinsic mechanism. Two examples that are probably
characteristics of a wide range of phenomena will be described in section III.
We then describe for comparison the salient features of the electron glass and
discuss the differences.

A modest conclusion that one may draw is that there are two types of glassy
electronic systems; those that can be unambiguously shown to be extrinsic, and
those that cannot, and for which the only consistent scenario is the electron
glass. However, we will argue that, on empirical basis, there is a positive
test for the electron glass; This is the memory dip that on one hand is
unequivocally associated with all the glassy features of the system, and on
the other hand its characteristic width is determined by the carrier
concentration of the material. The special features of the memory dip that are
suggestive of an electronic mechanism will be detailed and discussed.

\section{Experimental}

\subsection{Sample preparation and measurement techniques}

Several batches of samples were used in this study. These were thin films of
either polycrystalline or amorphous indium-oxide (to be referred to as
$In_{2}O_{3-x}$ and $In_{x}O$ respectively). The films were deposited by an
e-gun using 99.999\% pure $In_{2}O_{3}$. The amorphous films are typically
80-200~\AA ~thick, while the polycrystalline films have typically a thickness
of 30-60~\AA . As deposited indium-oxide films onto room temperature
substrates are amorphous. Polycrystalline films were prepared by heating the
sample to $\approx500~$K after deposition. The sheet resistance $R_{\square}$
of either structure was adjusted by thermal annealing (for $In_{x}O$) or
UV-treatment (for $In_{2}O_{3-x}$), to be within the range 2~M$\Omega
$-100~M$\Omega~$at 4K for the electron glass studies$.$ All samples had
lateral dimensions of $\simeq$1x1 mm. Some samples were configured as a
field-effect devices using either: 110$~\mu$m cover glass as spacers and a
gold film evaporated on the backside as gate, or 0.5$~\mu$m thermally grown
SiO$_{2}$ spacer on heavily doped Si wafer as gate. The 110$~\mu$m spacer is
used unless otherwise specifically mentioned. This is so as to enable the use
of reasonably high frequency $f$ in the conductance measurements even for
samples with very high resistance (note that the sample/gate capacitance is in
parallel with the ac measurement \cite{6}, and the RC problem limits the $f$
to low value thus compromising the temporal resolution and the signal/noise
value). The 0.5$~\mu$m spacer is used when it is desirable to cover a wide
range of charge variation in the field effect measurements without using an
excessively large voltage. Conductivity of the samples was measured using a
two terminal ac technique employing a 1211-ITHACO current preamplifier and a
PAR-124A lock-in amplifier for the 4K studies, and straightforward resistance
measurements using HP34401A for the room temperature measurements. Fuller
details of sample preparation, characterization, and measurements techniques
are given elsewhere \cite{6,11,18}. All resistance data reported here are per
square geometry (namely, data are specified in terms of the appropriate two
dimensional resistivity).

\section{Results and discussion}

Glassy features are readily observed in conductivity, and they may be
associated with various mechanisms. In this section we show and discuss some
experimental results of two groups of glassy effects revealed in conductance
measurements. Examples of glassy effects observable in the conductance that
can be shown to be due to extrinsic mechanisms are given in the next
subsection. This is followed by a list of features that should help to
distinguish the intrinsic electron glass from the extrinsic glass. In both
cases, the data used for illustration were deliberately taken on the same
systems namely, $In_{2}O_{3-x}$ and $In_{x}O$. The different glassy mechanisms
were achieved by varying the sample parameters (disorder) and the external
conditions used in the measurements (in particular, temperature).

\subsection{Glassy effects due to extrinsic mechanisms}

The first example we discuss in this category is the process of thermal
annealing, commonly used to reduce the resistance of amorphous indium-oxide
films, $In_{x}O$ \cite{18,19}. This involves holding the samples at a constant
temperature, typically up to few tens of degrees above room temperature, and
monitor the resistance versus time $R(t)$ as shown in Fig. 1 for a series of
samples, all made at the same deposition run. Note that $R(t),$ depicted here
on a semi-log scale, exhibits non-exponential relaxation. The observed slow
decrease in $R$ is an irreversible process and is due to elimination of
micro-voids which are common in amorphous materials prepared by quench cooling
\cite{20}. Naturally, this process also affects the density of the material,
namely, the sample volume shrinks as was directly confirmed by optical
measurements \cite{21}. This leads to a better wave-function overlap and thus
enhances the conductance. The slow relaxation of $R$ is then just a reflection
of the slow dynamics involved in the compaction process of the material. The
initial resistances of the samples used in Fig.~1 were of the order of
1-2~M$\Omega$, rather deep into the insulating regime making their conductance
very sensitive to even small changes of the landscape potential, and this is
the only reason for choosing them for the illustration. In fact, slow
conductance relaxations were observed even in samples that are well-annealed
to be on the metallic side of the transition \cite{21}.%
\begin{figure}
[ptb]
\begin{center}
\includegraphics[
trim=0.000000in 2.937617in 0.000000in 0.964472in,
height=3.122in,
width=3.3243in
]%
{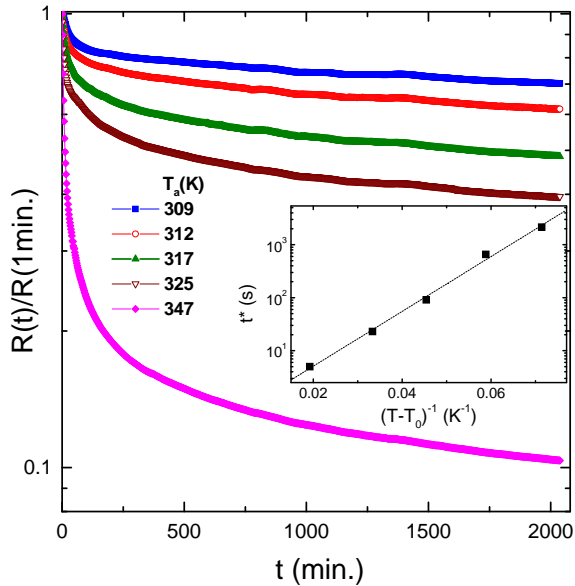}%
\caption{Resistance versus time-of-annealing (T$_{a}$ is the hot-stage
temperature) for a series of $In_{x}O$ samples each with initial sheet
resistance of $\approx2$ M$\Omega$. Data are normalized to the $R$ value
attained after 1~minute (where the hot-stage temperature reached $\approx95\%$
of its asymptotic value). Inset marks the time $t^{\ast}$ for the sample
resistance to drop to 70\% its value at 1 minute.}%
\end{center}
\end{figure}

Note that at any given time during the process $-\partial R(t)/\partial
t$\ increases rather fast with annealing temperature $T_{a}$, which varies by
only $\approx12$\% in the series. This presumably means that the annealing
process is activated, consistent with thermally assisted atoms diffusion. A
way to characterize the dynamics is to monitor the time $t^{\ast}$ it takes
for $R(t)$ to decay to a given fraction from its initial value. (This is,
admittedly, an arbitrary definition for a relaxation law that does not have a
characteristic scale but any other sensible definition will not change our
conclusions). In the inset to Fig.~1 we plot the dependence of $t^{\ast}$ on
temperature. As it turns out $t^{\ast}$ follows a temperature dependence of
the Vogel-Fultcher-Tammann type \cite{22}; $t^{\ast}\propto\exp\left[
\frac{E}{k_{B}(T-T_{0})}\right]  $, a typical behavior for many classical
glasses. The apparent divergence of $t^{\ast}$ at $T_{0}=295~$K (see inset to
Fig.~1) has a simple and plausible interpretation: The samples were prepared
and handled at or close to room temperatures prior to being subjected to the
annealing process, and thus they were already well annealed at $T_{0}$. No
glassy dynamics of this kind is observed with such samples at or below liquid
nitrogen temperatures.%
\begin{figure}
[ptb]
\begin{center}
\includegraphics[
trim=0.000000in 1.927811in 0.000000in 0.966739in,
height=3.5397in,
width=3.3243in
]%
{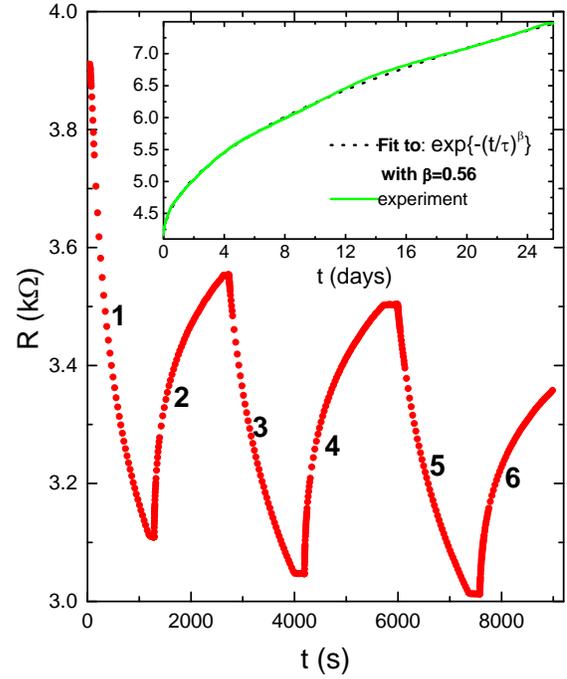}%
\caption{UV-treatment cycles of a $In_{2}O_{3-x}$ sample as reflected in its
sheet resistance $R$. During stages 1, 3, and 5 the sample is exposed to UV
irradiation ($\lambda=$280~nm) while in a vacuum chamber at pressure
p$\lesssim1~$mTorr. During stages 2, 4, and 6 the chamber is filled with air
to p$\approx$10$^{3}$~mTorr. Relatively flat parts occur when UV is turned off
(e.g., between stage 5 and 6), or when air is pumped away (e.g., between stage
4 and 5). Inset shows resistance versus time during an extended period of
exposure to air.}%
\end{center}
\end{figure}

Another set of glassy behavior observable through conductance measurement is
illustrated in Fig. 2. Here we utilize the flexibility of fine-tuning the
resistance of $In_{2}O_{3-x}$ films by UV-treatment. Unlike the thermal
annealing, this process is reversible, which is useful for many studies. While
under pumping, and constant UV illumination, the film resistance decreases
slowly. Breaking the vacuum by bleeding oxygen into the system reverses the
trend. Fig.~2 includes several cycles of UV-exposure+pumping, alternating with
bleeding-air periods resulting in a time sequence of $R(t)$ that obviously
retains some "memory" of previous conditions. A fit to a stretched exponential
can be easily obtained for each of the relaxations in Fig.~2. In fact, as
shown for example in the inset, for one of the `recovery' processes (which is
cheap to maintain for a long time), one could obtain an impressive fit to a
stretched exponential dependence extending over several weeks (more than 5
decades in time!)

History-dependence and stretched exponential behavior are sometimes taken as a
test of a "glass", so this example might have been considered a
\textit{bona-fide} indication of "electronic" glass. Note though that the
sample here is actually diffusive, and therefore it can not be an electron
glass, at least not from the point of view of the theories \cite{2,3,4,5}.
Actually, the reason for these slow conductance changes turns out to be
associated with change of stoichiometry in the system. A plausible scenario
that accounts for the glass-like dynamics in this case may be understood by
studying the structural changes caused by the UV-treatment process.

Indium-oxide films, whether amorphous or crystalline, are oxygen deficient
version of the ionic compound $In_{2}O_{3},$ and this non-stoichiometry gives
rise to `free' carriers (electron) density $n$. The further is $x$ in
$In_{2}O_{3-x},$ for example, the larger is $n.$ All other things being equal,
larger $n$ yields higher conductivity. This is essentially the same process as
when a semiconductor is doped by n-type impurities, except that the smallest
$n$ feasible in stable $In_{2}O_{3-x}$ films ($\simeq2\cdot10^{19}~$cm$^{-3})$
is already sufficient to make the system a degenerate Fermi gas. More
importantly, the process of "doping" here is easily affected by UV
illumination, which breaks the oxygen bonds, then pumping away the oxygen
atoms increases $x$ of the host. Exposing the sample to oxygen-rich atmosphere
reverses the process. Naturally, the processes of "doping" and "un-doping" are
reflected in the conductivity. The change of $n$ with the UV-treatment (and
the consequential mobility modification) were studied by a variety of ways
including Hall effect, and Burstein shifts using optical absorption \cite{23}.
Similar effects are observable in other materials such as $ZnO$ and they are
sometimes referred to as `persistent photo-conductivity' \cite{24} a term
originally reserved for electron-hole creation in semiconductors rather than
reflecting a change of the system chemistry \cite{25}. Another case of an
extrinsic effect that manifests itself as slow conductance relaxation is the
infamous "DX center" phenomenon. This is now believed to be a structural
effect well studied in semiconductors \cite{26}.

The reason for the sluggish response of the conductance are barriers that the
oxygen atoms have to overcome in crossing the film surface, \textit{and}
barriers for the atoms to diffuse into the bulk of the film. The dynamics of
these processes in general depends on a number of factors such as the
crystallite size, thickness of the film, and disorder (which makes the
relevant barriers a function of space, and thus distribute the diffusion rates
over a wide range. The phenomenon is also extremely sensitive to temperature;
For example, the hump in the curve between 12 to 16 days (inset to Fig.~2)
resulted from the temperature in the room increasing by $\approx2$ K (less
than 1\% change!)

Extrinsic glasses are interesting by their own right; the ease of monitoring
the underlying dynamics via the flexible conductance measurement has many
advantages for the study of the underlying glassy effects. The conductance in
these cases may then reflect all the basic glass features including aging and
other memory effects. However, such extrinsic glasses must not be confused
with the electron glass. There are some obvious differences between the
electron-glass and the extrinsic `conductivity-glasses' described above (and
many others of similar nature) such as temperature and its effect on dynamics,
and law of relaxation. It is desirable however to have a qualitative test that
distinguishes the inherent electron glass from extrinsic mechanisms. Such a
test is the existence of a memory dip, a feature that is probed by a field
effect measurement. It has phenomenology which is peculiar to the electron
glass, as will be detailed in sub-section c below.

\subsection{Intrinsic electron-glassy effects}

In theory, glassy effects in the conductance may originate from the interplay
between disorder and interactions while the ions are stationary. In fact,
sufficiently strong spatial disorder, such that the electronic states are
localized, is claimed to lead to a glassy state even in the non-interacting
system \cite{3} (Fermi glass). The presence of disorder however, has other
consequences; in the first place it gives rise to two-level-systems that, in
turn, manifest themselves in a wide spectrum of 1/f noise extending to very
low frequencies \cite{1}. This means that, even near equilibrium, ion/atom
dynamics occurs on time scales that, partially, overlaps with typical
relaxation times of the electron glass, as already remarked elsewhere
\cite{12}. Measurements on mesoscopic samples of $In_{2}O_{3-x}$ showed that
the electron glass dynamics occurs on a \textit{much} shorter time scale (and
different spatial scale) than the dynamics associated with structural TLS
suggesting that the two phenomena are of different origin \cite{15}. It seems
plausible that the two phenomena should influence one another, but there is
yet no evidence for such an effect. In particular, we failed to detect any
specific contribution to the 1/f noise due to the glassy state; for example,
no measurable change of noise could be detected when the sample was kept out
of equilibrium by a continuous excitation (created by continuously sweeping
the gate voltage).%
\begin{figure}
[ptb]
\begin{center}
\includegraphics[
trim=0.000000in 3.013551in 0.000000in 1.085739in,
height=3.039in,
width=3.3243in
]%
{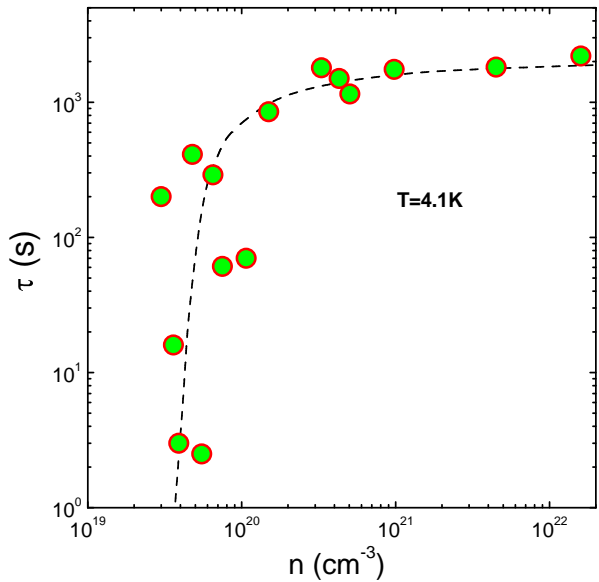}%
\caption{Typical relaxation time $\tau,$ as function of the carrier
concentration $n$ for a series of $In_{x}O$ samples (data are based on the
two-dip experiment, see \cite{17} and \cite{28} for fuller details and
interpretation). Note the sharp drop of $\tau$ for $n\lesssim10^{20}~$%
cm$^{-3}.$}%
\end{center}
\end{figure}

The other issue that the localization brings about is the impaired screening,
which makes the notion of a Fermi glass (non-interacting) of somewhat academic
scenario. Indeed, electron-electron interactions are believed to essential for
the existence of the glassy phase \cite{4,5}. Experimentally, the relevance of
interactions in real systems is evidenced by the observation that, all other
things being equal, the typical relaxation time $\tau$ of the electron glass
is controlled by the carrier concentration $n~$of the system \cite{17}$.$
Using a series of $In_{x}O$ films of similar resistance it was shown that
$\tau$ varies by 3 orders of magnitude for a change of a mere factor 4 in
$n.$~This is illustrated in Fig.~3 (based on data from \cite{17} and newer
data added recently using similar techniques). Other sample parameters such as
resistance, magnetic field, and temperature have a much smaller effect on
$\tau$~\cite{12,27,28}$.$ Finally, $\tau$ has been essentially unchanged when
the sample spatial extent has been varied from $\approx$1~cm down to 2$\mu m$
\cite{15}, and is also independent of thickness within the range studied.
Vaknin et al \cite{17} argued that these results imply the importance of
electron-electron interactions. It should be noted that the change in $n$ in
this series of samples amounts to changing the In/O ratio in the preparation
stage by a few percent. This hardly influences any structural aspects and
therefore it is hard to reconcile this observation with any `extrinsic'
mechanism. It should also be emphasized that these results are in agreement
with other methods of measuring the dynamics. These methods were fully
described in \cite{27,28}. The physics of this dramatic dependence on n shown
in Fig.~3, which empirically accounts for the absence of inherent slow
relaxation in low density systems (e.g., lightly-doped semiconductors as will
be further discussed below), has been interpreted as a quantum friction effect
in \cite{28}.

The other systems that show electron glassy effects (\textit{i.e.},
$In_{2}O_{3-x}$ and granular metals) lack the flexibility of varying $n$ over
a substantial range. ($n~$may be easily varied in semiconductors but their
$\tau~$appears to be quite short due to reasons explained elsewhere
\cite{17,28}). There is however, one feature that is common to all (inherent)
electron glasses - this is the memory dip (MD). The MD appears in field effect
measurements as a minimum of the conductance versus gate voltage $G(V_{g})$
centered around $V_{g}^{0}~-$ the gate voltage where the system was allowed to
equilibrate \cite{9,11,12}. It was initially termed `anomalous field effect'
by Ben-Chorin \textit{et al} where it was realized only in
`note-added-in-proof' that it is a non-equilibrium effect \cite{6}. As more
elaborate experiments became available, the intimate connection between this
`anomaly' and the electron glass properties became evident. In fact, every
single aspect of the electron glass, such as slow relaxation, memory effects,
and aging is reflected in the temporal behavior of the MD \cite{11}. For
example, it was shown that history dependence, such as aging, is wiped-out
when the MD is destroyed by either exposure to infrared source or non-ohmic
field \cite{29}. Also, for a given system, the electron glass dynamics, as
well as its temperature dependence, is uniquely characterized by the MD width
\cite{28}.%
\begin{figure}
[ptb]
\begin{center}
\includegraphics[
trim=0.000000in 0.867005in 0.000000in 0.726471in,
height=4.0836in,
width=3.3243in
]%
{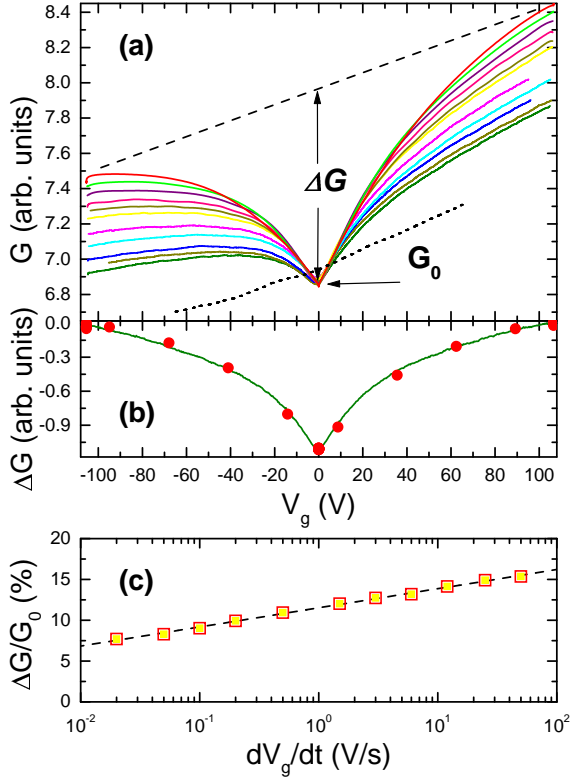}%
\caption{(a) Field effect as function of the gate-voltage sweep-rate, showing
11 different rates (with values depicted in 4(c)). Sample is $In_{2}O_{3-x}$
with $R=$26~M$\Omega$ at 4~K. Dashed line indicates linear approximation to
the equilibrium field effect for the fastest rate. Dotted line is an actual
result of a sweep rate during two days over the range shown. (b) Illustrates
the collapse of the MD data for the slowest and fastest sweep rates in (a)
after subtracting the linear part of the field effect and normalizing the
amplitude. (c) The MD normalized amplitude $\Delta G/G$ as function of sweep
rate illustrating the basic log(t) dependence.}%
\end{center}
\end{figure}

\subsection{The memory dip (MD) properties}

The physics underlying the function $G(V_{g})$ we refer to as MD is not yet
well understood. Several authors have conjectured that the MD is a reflection
of the Coulomb gap \cite{30,31,32,33}. Adopting this view indeed helps to
understand some of the non-trivial features exhibited by $G(V_{g})$ of
electron-glasses, such as the two-dip experiment \cite{10,12}. On the other
hand, some difficulties with this interpretation were also pointed out
\cite{9,17}. In the following, these reservations will be re-examined in view
of recent experiments.

We start by reviewing the basic properties of the phenomenon at a given
temperature, as the great majority of experiments were performed at or around
$\approx$4K. The MD revealed in the $G(V_{g})$ scans is characterized by its
relative amplitude $\Delta G/G$ and by its shape (functional form). There are
several variables that may affect appreciably $\Delta G/G,$ however, the
functional form of the MD is \textit{independent} of \textit{any} of these
variables. For example, increasing the sweep rate of the gate voltage $V_{g},$
increases $\Delta G/G$ (logarithmically, see, Fig.~4a,c) yet, upon re-scaling
\textit{just} \textit{the y-axis}, the MD of two scans with quite different
$\partial V_{g}/\partial t$ would perfectly overlap one another (Fig.~4b). The
dependence on $V_{g}$ sweep-rate (that in this work has been extended to cover
more than 3 decades) also illustrates that the MD is a non-equilibrium
phenomenon as well as the basic $\log(t)$ dynamics.%
\begin{figure}
[ptb]
\begin{center}
\includegraphics[
trim=0.000000in 0.964472in 0.000000in 0.726471in,
height=4.0413in,
width=3.3243in
]%
{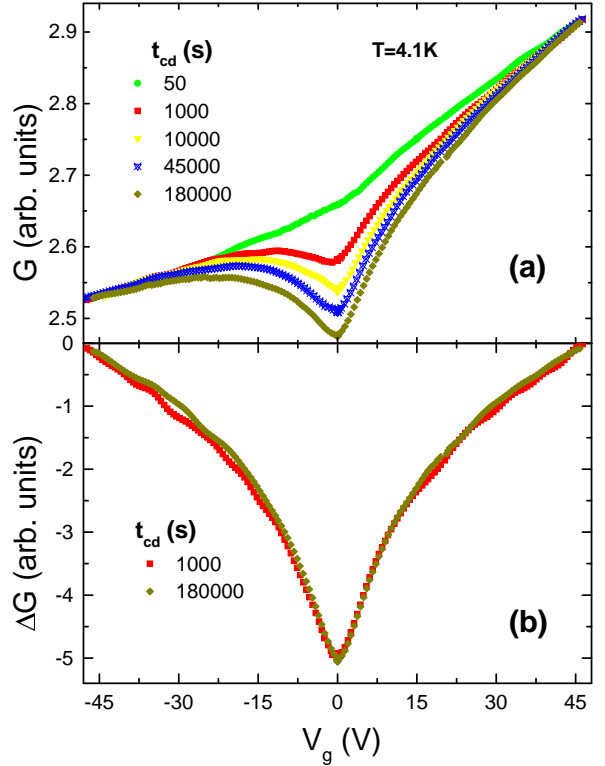}%
\caption{(a) Field effect sweeps taken at several times since cool-down
t$_{cd}$ showing the MD slow evolution. Sample is $In_{2}O_{3-x}$ film with
$R=49~$M$\Omega$. (b) Shows the collapse of the MD shape for the data for the
sweeps taken at t$_{cd}=10^{3}$ seconds and t$_{cd}=180,000$ seconds (after
subtracting the equilibrium field effect and expanding the t$_{cd}=10^{3}$
seconds MD data by a constant factor).}%
\end{center}
\end{figure}

Another factor that affects the amplitude but not the shape of the MD is the
time during which the system equilibrates after the initial cool-down to the
measurement temperature (Fig.~5). The evolution of $-\Delta G$ with time at
$V_{g}$=0 is just the $\log(t)$ law of the electron glass that one observes in
$G(t)$, say after a quench from high temperature (even without configuring the
sample with a gate). The interesting thing here is that while this conductance
`equilibration point' is slowly going down it "drags" with it some region
around $V_{g}$=0 in such a manner to keep the shape of the symmetric field
effect (\textit{i.e.}, the MD) constant (\textit{c.f}., inset to Fig.~5). Note
that this region extends over a rather large range of $V_{g},$ a point to
which we shall return below.

Finally, two other factors that do not affect the MD shape; Disorder and
magnetic field. By higher disorder we refer to samples that are deeper into
the insulating phase. This is manifested by a larger $R~$at a given $T.~$Both,
the slope of asymmetric field effect and $~\Delta G/G$ of the MD increase with
the sample resistance. However, disorder has no effect on the shape of the MD
\cite{11}. A magnetic field as high as 30T, while changing $G$ by as much as a
factor \cite{34} of 4, which naturally also affects the equilibrium
(asymmetric) part of the field effect, does not change the MD shape (Fig.~6).
Interestingly, the magnetic field changes both $\Delta G~$and~$G$ in such a
way that $\Delta G/G$ is constant. Other agents that modify $G$, also change
$\Delta G/G.$ The reason for that is not currently understood.

The only factor that, at a fixed temperature, does affect the MD shape is the
carrier concentration $n~$of the system. The characteristic width of the MD
(to be defined below) increases monotonically with $n,$ \textit{and that seems
to be a `universal' feature, common to all electron glasses studied to date}
$In_{2}O_{3-x},$ $In_{x}O,$ and granular metals \cite{15}. Note that this
observation clearly implicates an electronic mechanism, and qualitatively it
is in line with the Efros-Shklovskii Coulomb gap \cite{35,36}, which increases
monotonically with the density of states \cite{36}. As mentioned in section I,
the possibility that a Coulomb gap (or, more generally, an interaction gap) is
the underlying physics of the MD was considered, but several reservations were
raised against it. A serious concern is the relevance of a single-particle
density-of-states (DOS) to conductance. In addition, a severe problem was a
factor of $\gtrsim10$ discrepancy \cite{17}\ between the experimental results
for the width of the MD and the value expected for the strength of the Coulomb
interaction (\cite{17} used $\approx\frac{e^{2}}{\kappa r}$, with a dielectric
constant $\kappa\approx10$ and $r\approx n^{-1/3}$ as a measure of this energy).

It is this discrepancy that we wish to re-examine here and show that it was
based on a definition of width that, though natural, may have underestimated
its value.%
\begin{figure}
[ptb]
\begin{center}
\includegraphics[
trim=0.000000in 0.843205in 0.000000in 0.605203in,
height=4.1425in,
width=3.3243in
]%
{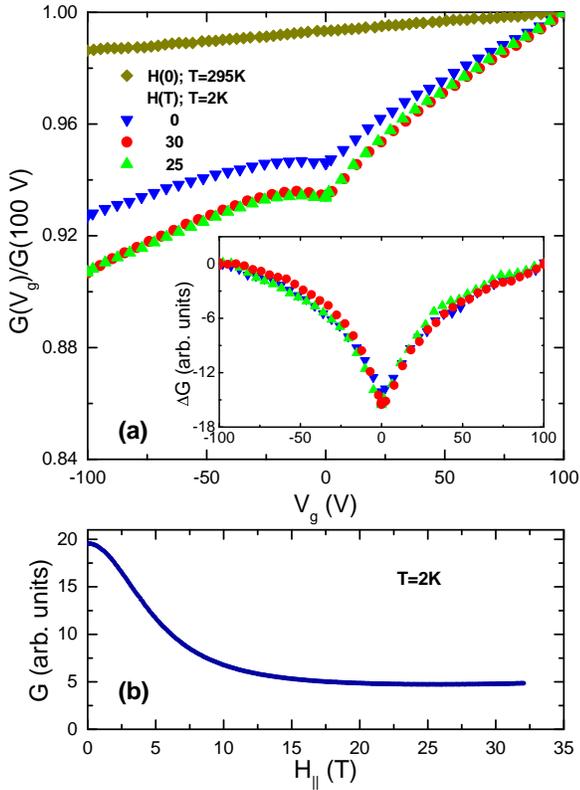}%
\caption{(a) Field effect sweeps taken for different values of a magnetic
field (parallel to a $In_{2}O_{3-x}$ film plane). The sample is 32~\AA
\ thick, and has sheet resistance of 500~k$\Omega$ at 4.2~K and 2.11~M$\Omega$
at 2~K. For comparison, the field effect of this sample at room temperature is
also shown. Note the change of the equilibrium field effect with either,
temperature, and magnetic field (the sample resistance increases with magnetic
field; see lower graph). Inset shows the preservation of the MD shape and
magnitude (in this case just the linear field effect has been subtracted from
the $G(V_{g})$ data). (b) The magneto-conductance of the film. For larger
fields, quantum interference effects (which exhibit positive
magneto-conductance), start to take over the spin effects. This creates the
minimum in $G(H)$ at 25~T. }%
\end{center}
\end{figure}

The definition of the MD width adopted by Vaknin \textit{et al }%
\cite{17}\textit{ }is based on measuring the width (in volts) of the MD at
half-height from the $G(V_{g})$ scan. Using the gate-sample capacitance the
width in terms of the associated charge $\Delta Q$ is obtained \ Note that it
is this charge range $\Delta Q$ rather than the voltage span $\Delta V_{g}$
which is the relevant variable to characterize the width of the MD. This has
been demonstrated by comparing FET samples with different spacers and spacer
thickness for a given electron glass \cite{11}. Then, the width in energy
units is derived with the $\Delta Q$ associated with the half-height and the
calculated density of states using the measured $n$ and free electron
formulae.%
\begin{figure}
[ptb]
\begin{center}
\includegraphics[
trim=0.000000in 1.012073in 0.000000in 0.506603in,
height=4.1139in,
width=3.3243in
]%
{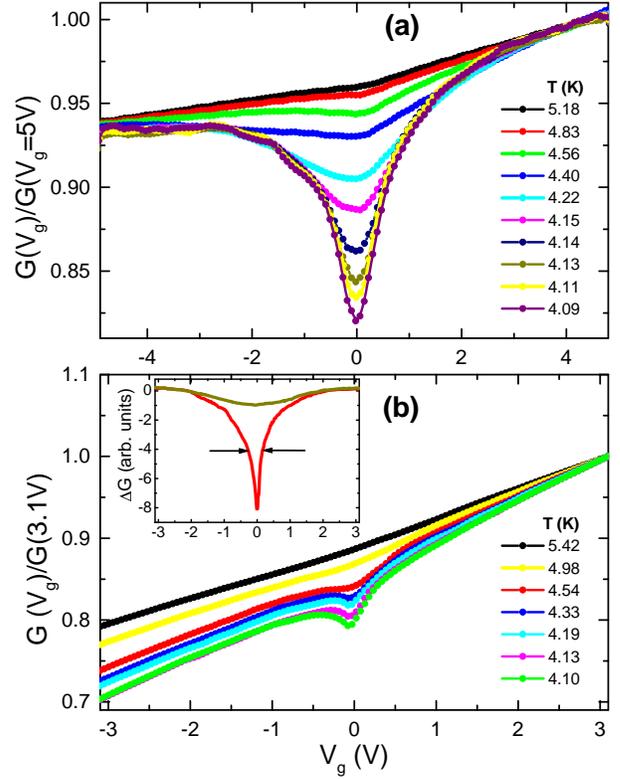}%
\caption{Dependence of the field effect on temperature for $In_{x}O$ sample
(a) with $R=25.2~$M$\Omega$ at 4~K, and $In_{2}O_{3-x}$ sample (b) with
$R=17.6~$M$\Omega~$at 4~K. In both cases the spacer is 0.5~$\mu$m of thermally
grown SiO$_{2}$ on heavily doped Si wafer. Inset to (b) is the MD at 4.1~K and
5.42~K (after a linear part of the equilibrium field effect has been
subtracted from the $G(V_{g})$ data).}%
\end{center}
\end{figure}

Since the half-height of the MD (to be referred to as $\Gamma_{1/2})$ is
experimentally well-defined, this procedure of assigning a width to a
particular sample is convenient and unambiguous. However, if the modulation in
$G(V_{g}),$ or more precisely, $G(Q)$, reflects a DOS versus energy structure,
then the shape of the MD would depend on temperature due to the thermal
occupation of states. This indeed seems to be the case; the MD gets broader as
$T$ increases \cite{37}. The price for the experimentally convenient
definition is a spurious temperature dependent energy scale. In simple words,
the temperature dependence of $\Gamma_{1/2}$ is an artifact that arises from
the non-linear mapping between added charge, and the associated energy change.
This point was elaborated on by Lebanon and Mueller that, using a model,
showed that the MD shape becomes more "cusp-like" as the temperature is
reduced is a way that qualitatively resembles the experiments \cite{33}%
\ (however, the specific relation between the assumed DOS and $G$ is still to
be justified). Using the `half-height' $\Gamma_{1/2}$ criterion for comparison
with the Coulomb gap width is therefore problematic, which, in hindsight,
being \thinspace temperature dependent was was not relevant for the comparison
to start with.

An alternative way to define the energy scale associated with the MD suggests
itself when one studies the temporal evolution of $G(V_{g})$ after a
quench-cool from high temperature as shown in Fig.~5 above. Note that the
$G(V_{g})$ scans of this sample show time dependence throughout the interval
(-50~V to +50~V) while outside it $G$ tend to approach a time independent
regime. The precise boundaries of this region may be ill defined, yet
obviously, the `active' energy band extends over a wide range that does not
depend on time or on $\Delta G/G$. It makes sense then to use this `active'
region as the relevant $V_{g}$ scale and derive the MD width (henceforth
labelled $\Gamma)$ from it. Indeed, this definition for the MD width turns out
also to be temperature independent. This is illustrated in Fig.~7a for a
$In_{x}O$ sample, and in Fig.~7b for $In_{2}O_{3-x}$. (Note that in these
experiments, the equilibrium (asymmetric) part of the field effect changes
with temperature. This results from the the higher sensitivity of $\Delta
G/\Delta Q$ as the temperature is lowered and $R$ increases (compare with
Fig.~5 where $T$ is fixed). Note incidentally that the slope of the
equilibrium field effect in $In_{x}O$ is considerably smaller than that of the
$In_{2}O_{3-x}$ sample (Fig.~7). This is a result of the shallower slope of
the thermodynamic DOS versus energy in the amorphous material.

The pronounced temperature dependence of $\Gamma_{1/2}$ of these samples are
shown in Fig.~8a. For comparison, the dependence of the half-height width on
temperature for a $In_{2}O_{3-x}$ FET sample with a much thicker spacer is
shown in Fig.~8b, which shows that $\Gamma_{1/2}(V)$ at a given $T$ does scale
with the thickness of the spacer (which affects the sample-gate capacitance)
as mentioned above.%
\begin{figure}
[ptb]
\begin{center}
\includegraphics[
trim=0.000000in 3.303686in 0.000000in 1.254607in,
height=2.8478in,
width=3.3243in
]%
{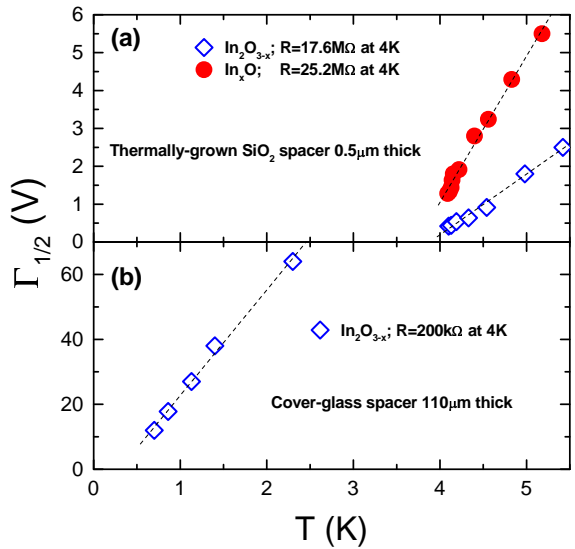}%
\caption{The dependence of the half-height width $\Gamma_{1/2}$ on
temperatures for the sample in Fig.~7 and for a $In_{2}O_{3-x}$ sample
configured with a 110~$\mu$m of SiO$_{2}$ [(a) and (b) respectively].}%
\end{center}
\end{figure}

The ratio $\frac{\Gamma}{\Gamma_{1/2}}\ ($where $\Gamma_{1/2}~$is taken at
$T\approx$4K), is $\approx8$ for the $In_{x}O$ in Fig.~7a, and $\approx10$ for
the $In_{2}O_{3-x}$ sample in Fig.~7b, and a ratio of 7-10 was found for the
series of $In_{x}O$ samples with $n$ in the range 6$\cdot10^{19}$ cm$^{-3}$ to
8$\cdot10^{21}$ cm$^{-3}$. In terms of energy, this range amounts to 10-80~meV
respectively, a range that compares favorably with the associated electronic
energy $\approx\frac{e^{2}}{\kappa}n^{1/3}.$ These values for the MD width are
not too far from the values expected for the Efros-Shklovskii Coulomb gap,
leads us to re-consider the possible connection between the two.

The main difficulty in trying to compare the MD with the Coulomb gap is
related to the different prescriptions for measuring them. The Coulomb gap is
a single-particle DOS which, by definition, is associated with adding charge
to the system while all other particles are frozen (like, \textit{e.g}., in
tunneling or photo-emission experiment). It is also an equilibrium entity.
This is not the situation in the field effect measurements where, for any
feasible sweep rate, some relaxation takes place, and when carried out under
equilibrium conditions no MD is observed (\textit{c.f}. the dotted line in
Fig.~4a). Nevertheless the two phenomena seem to be linked to a similar energy
scale. Clarifying the possible relation between the Coulomb gap and the MD is
a challenge to theory.

A related issue is how the `interaction-band' associated with the MD affects
the equilibrium conductance $G.$ Experimentally, the conductance versus
temperature $G(T)$ in all electron glasses is always activated, as it must be
for a system with localized states. However, no characteristic $G(T)$ law can
be identified. For example, $In_{2}O_{3-x}$ films usually exhibit
$G(T)\propto\exp\left[  -(\frac{T_{0}}{T})^{\alpha}\right]  $ with
$\alpha\approx1/3$ when deeply insulating (namely, Mott-like behavior in a 2D
system) and $\alpha\approx1$ when closer to the diffusive regime. Granular
metals and $In_{x}O$ \ films often show stretched exponential $G(T)$ with
$\alpha\approx1/2$ while in granular Al films $\alpha\approx0.8~$has been
reported \cite{8}$.$ It appears that there is more diversity in the $G(T)$
laws of the electron glasses than in their non-equilibrium properties, which
are remarkably similar. Therefore, slow conductance relaxation due to the
electron glass scenario may be positively identified by the behavior of the
associated MD but not by\ $G(T)$.

\subsection{Discussion}

The experimental observations detailed above raise the question: What about
those systems that do show hopping conductivity yet do not show MD in field
effect measurements; are they not electron glasses? This is quite a pertinent
question; there are many such systems, notably semiconductors. To our best
knowledge, a memory dip has not been reported in any semiconductor even at
temperatures lower than 1~K.

Theoretically, the electron glass is generic to a degenerate Fermi system with
localized states and Coulomb interactions. The theory conjectures that all
such systems are electron glasses (and, by our proposed test, should exhibit
MD in field effect measurements). Theory however does not specify the typical
relaxation time of a given system, so here again we have to rely on
experiment. All experiments performed to date suggest that the relaxation time
$\tau$, and thus the time for the MD to `form' and `decay', is a monotonous
function of $n$ and relaxation times that are appreciably larger than few
seconds are found only in systems with $n\gtrsim10^{20}$~cm$^{-3}.$

As is shown in Fig.~3, below a certain carrier concentration $n$, $\tau$ drops
down \textit{extremely} fast. Such a `critical' concentration may vary to some
degree between different systems, yet it is hard to see what else except $n$
(and to a considerably lesser degree $R$) may have an appreciable effect on
$\tau$ for a generic effect that hinges on just localized states and Coulomb
interactions \cite{38}. A possible reason for the dramatic $n$ dependence
based on quantum friction scenario was offered in \cite{28} but here we shall
treat this finding on purely empirical grounds. Since semiconductors that are
in the hopping regime have typically $n\ll10^{19}~$cm$^{-3}$ it is quite
conceivable that their typical $\tau$ will be much smaller than 1~second, as
suggested by Fig.~3. That will make it very hard to see the MD in a field
effect experiment, which is an inherently slow process due to the gate-sample
capacitance and sample resistance. Simply stated, a consistent answer is that
\textit{all} hopping systems are in principle electron glasses provided their
resistance is large enough, and they are measured at low enough temperature.
However, if no MD is observed it might mean that their `slow' relaxation is
too short to be observed in the field effect. In this situation, observing
conductance relaxation for larger time than that which would allow MD
detection should be a warning sign of an extrinsic mechanism.

In summary, we discussed the differences between extrinsic and intrinsic
glassy effects observed in conductance measurements. A simple test for an
intrinsic electron glass has been proposed based on a field effect measurement
performed on the system after allowing it to equilibrate. The appearance of a
dip with dynamics that precisely coincides with the relaxation of the excess
conductance caused by, e.g., a quench from high temperatures, is a primary
signature of the electron glass. In addition, the dip should have a
characteristic shape that does not depend on disorder, magnetic field, or
gate-voltage sweep-rates, and width that is of order of a relevant electronic
energy. Moreover, the shape of the dip should depend on temperature even for
$T\ll\Gamma$; an attribute that is peculiar\ to the electron glass, quite
distinct from thermal smearing.

It would be of great interest to extend the study of the electron glass to
include more materials, and in particular, to further test the empirical
connection between carrier concentration and relaxation time. Systems that are
may be promising in this regard include high-Tc compounds, and transition
metal semiconductor alloys, provided that they are sufficiently deep into the
hopping regime. On the theoretical side, there is the challenge of dealing
with the non-equilibrium DOS that is associated with the memory dip, which
seems to be the essence of the electron glass physics.

The author acknowledges with gratitude the hospitality and use of facilities
at the NHMFL at Tallahassee, and illuminating discussions with A. L. Efros.
This research was supported by a grant administered by the US Israel
Binational Science Foundation and by the Israeli Foundation for Sciences and Humanities.


\begin{thebibliography}{99}                                                                                               %


\bibitem {1}P. Dutta, and P. M. Horn, Rev. Mod. Phys. \textbf{53}, 497 (1981);
M. B. Weissman, Rev. Mod. Phys., \textbf{60}, 537 (1988), and references therein.

\bibitem {2}M. Gr\"{u}newald, B. Pohlman, L. Schweitzer, and D. W\"{u}rtz, J.
Phys. C, \textbf{15}, L1153 (1982).

\bibitem {3}M. Pollak and M. Ortu\~{n}o, Sol. Energy Mater., \textbf{8}, 81
(1982); M. Pollak, Phil. Mag. B\textbf{ 50}, 265 (1984).

\bibitem {4}J. H. Davies, P. A. Lee, and T. M. Rice, Phys. Rev. Lett.,
\textbf{49}, 758 (1982).

\bibitem {5}G. Vignale, Phys. Rev. B\textbf{ 36}, 8192 (1987).

\bibitem {6}M. Ben-Chorin, D. Kowal and Z. Ovadyahu, Phys. Rev. B \textbf{44},
3420 (1991).

\bibitem {7}G. Martinez-Arizala, D. E. Grupp, C. Christiansen, A. Mack, N.
Markovic, Y. Seguchi, and A. M. Goldman, Phys. Rev. Lett., \textbf{78}, 1130
(1997). G. Martinez-Arizala, C. Christiansen, D. E. Grupp, N. Markovic, A.
Mack, and A. M. Goldman, Phys. Rev. B\textbf{ 57}, R670 (1998).

\bibitem {8}T. Grenet, Eur. Phys. J, \textbf{32}, 275 (2003); T. Grenet, J.
Delahaye, M. Sabra, and F. Gay, Eur. Phys. J. B \textbf{56}, 183 (2007).

\bibitem {9}M. Ben Chorin, Z. Ovadyahu and M. Pollak, Phys. Rev. B\textbf{
48}, 15025 (1993).

\bibitem {10}A. Vaknin, Z. Ovadyhau, and M. Pollak, Phys. Rev. Lett.,
\textbf{84}, 3402 (2000).

\bibitem {11}A. Vaknin, Z. Ovadyahu, and M. Pollak, Phys. Rev. B\textbf{ 65},
134208 (2002).

\bibitem {12}M. Ben Chorin, Z. Ovadyahu and M. Pollak, Phys. Rev. B\textbf{
48}, 15025 (1993); Z. Ovadyahu and M. Pollak, Phys. Rev. Lett., \textbf{79},
459 (1997).

\bibitem {13}Ariel Amir, Yuval Oreg, and Yoseph Imry, Phys. Rev. B 77, 165207
(2008); A. Vaknin, Z. Ovadyahu, and M. Pollak, Phys. Rev. B \textbf{61}, 6692 (2000).

\bibitem {14}Z. Ovadyahu, and M. Pollak, Phys. Rev. B\textbf{ 68}, 184204 (2003).

\bibitem {15}V. Orlyanchik, and Z. Ovadyahu, Phys. Rev. B \textbf{75}, 174205 (2007).

\bibitem {16}T. Grenet, Phys. Stat. Sol. C \textbf{1}, 9 (2004); J. Delahaye,
T. Grenet and F. Gay, Eur. Phys. J. B 65, 5 (2008).

\bibitem {17}A. Vaknin, Z. Ovadyhau, and M. Pollak, Phys. Rev. Lett.,
\textbf{81}, 669 (1998).

\bibitem {18}Z. Ovadyahu, J. Phys. C: Solid State Phys., \textbf{19}, 5187 (1986).

\bibitem {19}A. T. Fiory, and A. F. Hebard, Phys. Rev. Lett. \textbf{52}, 2057
(1984); G. Sambandamurthy, L.W. Engel, A. Johansson, and D. Shahar, Phys. Rev.
Lett., \textbf{92}, 107005 (2004).

\bibitem {20}N. F. Mott and E. A. Davis, in: "\textit{Electronic Processes in
Non-Crystalline Materials}", pp 336-338, Clarendon Press, Oxford (1979).

\bibitem {21}Z. Ovadyahu, Phys. Rev B \textbf{47}, 6161 (1993).

\bibitem {22}Jeppe C. Dyre, Rev. Mod. Phys. \textbf{78}, 953 (2006) and
references therein.

\bibitem {23}O. Cohen and Z. Ovadyahu, Phys. Rev. B \textbf{50}, 10442 (1994).

\bibitem {24}Z. Y. Xiao, Y. C. Liu, R. Mu, D. X. Zhao, and J. Y. Zhang, Appl.
Phys. Lett. \textbf{92}, 052106 (2008).

\bibitem {25}I. Giaever and H. R. Zeller, Journal of Vacuum Science and
Technology, \textbf{6}, 502 (1969); P. Stradins and H. Fritzsche, Phil. Mag. B
\textbf{69}, 121 (1994).

\bibitem {26}See e.g., Jingbo Li, Su-Huai Wei, and Lin-Wang Wang, Phys. Rev.
Lett., 94, 185501 (2005); S. Lany , H.Wolf , and T. Wichert, Phys. Rev. Lett.,
92, 225504 (2004).

\bibitem {27}Z. Ovadyahu, Phys. Rev. B \textbf{73}, 214208 (2006).

\bibitem {28}Z. Ovadyahu, Phys. Rev. Lett., \textbf{99}, 226603 (2007).

\bibitem {29}V. Orlyanchik, A.Vaknin, Z. Ovadyahu, and M. Pollak, Phys. Stat.
Sol., \textbf{b230}, 61 (2002); V. Orlyanchik and Z. Ovadyahu, Phys. Rev.
Lett., 92, 066801 (2004).

\bibitem {30}C. C. Yu, Phys. Rev. Lett., \textbf{82}, 4074 (1999); Physica
Status Solidi \textbf{b230}, 47 (2002).

\bibitem {31}D. R. Grempel, Europhys. Lett., \textbf{66,} 854 (2004); A. B.
Kolton, D. R. Grempel, and D. Dominguez, Phys. Rev. B\textbf{\ 71}, 024206 (2005).

\bibitem {32}M. M\"{u}ller and L. B. Ioffe, Phys. Rev. Lett. \textbf{93},
256403 (2004).

\bibitem {33}Eran Lebanon, and Markus Mueller, Phys. Rev. B\textbf{\ 72},
174202 (2005).

\bibitem {34}The positive magnetoresistance results from the spin alignment
mechanism proposed by: A. Kurobe and H. Kamimura, J. Phys. Soc. Japan,
\textbf{51}, 1904 (1982).

\bibitem {35}M. Pollak, Discuss. Faraday Soc. \textbf{50}, 13 (1970).

\bibitem {36}A. L. Efros and B. I. Shklovskii, J. Phys. C: Solid State Phys.,
8, \textbf{L49} (1975).

\bibitem {37}A. Vaknin Z. Ovadyahu, and M. Pollak, Europhys. Letters
\textbf{42}, 307 (1998).

\bibitem {38}This however is based on experiments in the liquid helium
temperature range where $\tau$ is only weakly temperature dependent. It
remains to be seen whether long relaxation times are achievable even in
systems with low carrier concentration when cooled to ultra low temperatures
(while still of measurable resistance).
\end{thebibliography}
\end{document}